\documentclass{article}
\usepackage[utf8]{inputenc}
\usepackage[english]{babel}
\usepackage{newtxtext,newtxmath}
\usepackage[dvipdfmx]{graphicx}
\usepackage{authblk}
\usepackage{url}

\begin{document}
\title{Inference of Personal Attributes from Tweets Using Machine Learning}

\author[$\dagger$]{Take Yo}
\author[$\dagger$,$\ddagger$]{Kazutoshi Sasahara\thanks{Correspondence should be addressed to K.S. (sasahara@nagoya-u.jp)}}
\affil[$\dagger$]{Graduate School of Informatics, Nagoya University, Furo-cho, Chikusa-ku, Nagoya 458-8601, Japan}
\affil[$\ddagger$]{JST, PRESTO, Kawaguchi, Japan}
\date{}

\maketitle

\begin{abstract}
Using machine learning algorithms, including deep learning, we studied the prediction of personal attributes from the text of tweets, such as gender, occupation, and age groups.
We applied word2vec to construct word vectors, which were then used to vectorize tweet blocks. 
The resulting tweet vectors were used as inputs for training models, and the prediction accuracy of those models was examined as a function of the dimension of the tweet vectors and the size of the tweet blocks. 
The results showed that the machine learning algorithms could predict the three personal attributes of interest with 60--70\% accuracy.
\end{abstract}


\section{Introduction}
Social media was developed for people who desire diverse communication paradigms. 
Currently, social media plays the role of a hub for social information, a platform for exchanging opinions, and a place for researchers to observe digital traces of human behavior. 
In addition to the availability of big data and improvements in computing capability with GPGPU, machine learning techniques, including deep learning, are more practical for image, sound, and natural language processing. They are becoming important tools for promoting the social sciences and constructing social information infrastructures~\cite{Huang2012,Lin2014,Lin2014a}.
Based on this background, a new interdisciplinary science called computational social science has emerged and it has been actively investigated in recent years~\cite{Lazer2009,Golder2014}.

According to the official announcement from Twitter, as of April 2016, there are approximately 310 million monthly active accounts worldwide, and the number of monthly views of tweets with embedded photos and videos has reached 1 billion~\cite{twitter1}. 
Although the increase in the number of active accounts is slowing year after year for social media competitors (such as Facebook and Instagram), there are about 40 million monthly active Twitter accounts in Japan, as of September 2016. 
This indicates that Twitter is still a popular social media platform in Japan~\cite{twitter2}. 
Unlike Facebook, which requires permissions for social networking, one can follow both friends and others of interest without permissions. 
After that, friends can share any contents online. 
This follow mechanism enables ``loose social relationships'' that encourage diverse communications. 
Besides this, Twitter allows us to collect a large amount of social data via API~\cite{twitter_api}. 
Thus, Twitter has frequently been used as a data source by computational social science researchers (e.g., \cite{Golder:2011cy,Sasahara:2013eu,Takeichi:2015ia,Kaur2016}). 

The prediction of personal attributes based on social data has become a major research theme in recent years. 
Previous research has examined the relationships between personal attributes and behaviors. 
For example, how people talk and write are known to be associated with various personal attributes, such as educational background and growth environment~\cite{Pennebaker2007,Tausczik2010}. 
However, previous research has been limited by the availability of data.
In the age of social media, people spontaneously post and share linguistic expressions online, and this information can be used to infer personal attributes~\cite{Golder2014}. 
Although a significant amount of research has been done on this topic, little is known about how to computationally infer personal attributes from social data, and no versatile algorithm has yet been established. 

The aim of this research is to investigate the extent to which personal attributes can be predicted only from texts in social media posts.
This results can give us baseline data; we can further increase the prediction accuracy by adding other features in addition to texts.
In this study, we used a large dataset from Twitter, transferred tweets to vectors using a word embedding method, and then predicted gender (male or female), occupation (10 different jobs), and age groups (whether he/she was born before 1980, indicating ``digital native'' or ``digital immigrant'') based on those tweet vectors using five different machine learning algorithms. 
In this paper, we report two preliminary results.

\begin{figure}[t]
\begin{center}
 \includegraphics[clip,width=1.0\textwidth]{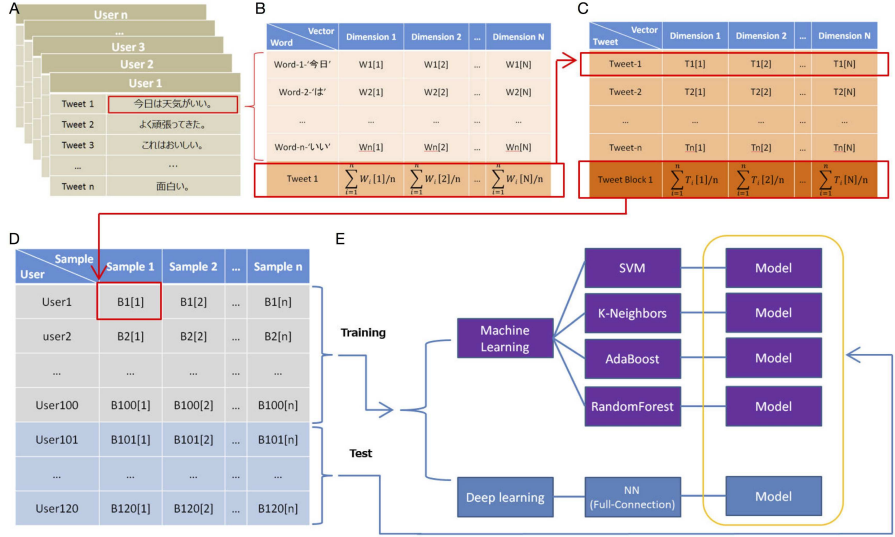}
 \caption{Schematic diagram of data processing and machine learning}
 \label{pic1}
\end{center}
\end{figure}

\section{Related Research}
With the development of information technology, the inference of personal attributes based on social data has been actively studied for applicability. 

Sloan et al. showed that demographic information (gender, language, location, age, occupation, and social class) could be accurately extracted from the profile descriptions of Twitter users using natural language processing (NLP)~\cite{Sloan2013, Sloan2015}.
Schwartz et al. analyzed words, phrases, and topics from Facebook posts. 
Combined with personality tests, they observed close relationships among language use and personality, gender, and age ~\cite{Schwartz2013}. 
Kosinski et al. demonstrated that even simple algorithms can predict personal attributes on the bias of the patterns of Facebook's ``likes,'' an indicator of peoples' preferences~\cite{Kosinski2013}. 
Wang et al. applied various deep learning algorithms to extract information from tweets, profile images and posted pictures and predicted their political intonation~\cite{Wang2017}.
Liu and Zhu demonstrated that the Big Five factors in human personality (i.e., openness, conscientiousness, extraversion, agreeableness, and neuroticism) could be predicted from text posts on the Chinese microblogging platform Weibo~\cite{Liu2016}. 
IBM has also developed a service called Personality Insights which predicts personality traits including the Big Five factors, needs, and values~\cite{IBM}. 

The inference of personal attributes from social data is increasingly studied in academia and industry because it can be applied to a wide range of areas, including basic research in social science and applications for information recommendation and social media marketing.

\begin{figure}[t]
\begin{center}
 \includegraphics[clip,width=1.0\textwidth]{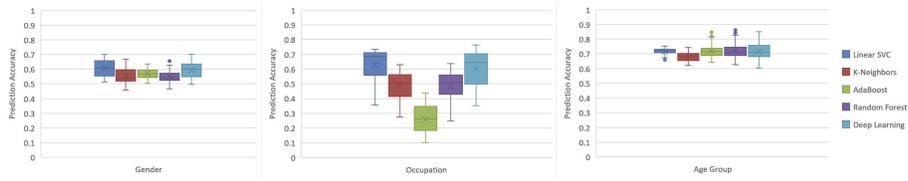}
 \caption{Comparison of different algorithms in prediction accuracy for three attributes}
 \label{pic2}
\end{center}
\end{figure}

\section{Method}
\subsection{Data Collection}
To construct models that predict gender, occupation, and age groups, we collected tweets from Japanese accounts via. Twitter API~\cite{twitter_api}. 
We manually selected 120 active Twitter accounts whose posts numbered more than 3000 at the time of crawling and for which genders and occupations could be identified from the personal Twitter profiles or reliable information sources, such as Wikipedia. 
We then assigned age group labels to these accounts using reliable information sources as much as possible, because age information is less readily available compared to gender and occupation. 
Among the 120 accounts, we used tweets from 100 accounts for training and tweets from the remaining 20 accounts for testing.

In the 120 accounts, the number of males and females was the same. The number of accounts belonging to each of 10 types of jobs (politician, entertainer, cartoonist, entrepreneur, scholar, journalist, writer, musician, athlete, IT engineer) was also the same. 
The age groups were divided into ``digital natives'' born after 1980, ``digital immigrants'' born before 1980, and ``unknowns'' who were not identified via internet search. 
In the training data from 100 accounts, there were 50 digital natives, 41 digital immigrants, and 9 unknowns. 
In the test data from 20 accounts, there were 11 digital natives, and 9 digital immigrants. 
Thus, the number of people in age group was approximately the same. 
Next, we crawled user timelines, which included retweets and replies from each account as much as possible (Fig.~\ref{pic1}A). 
This resulted in 314,382 tweets from the training accounts and 64,027 tweets from the test accounts.

\subsection{Data Processing}
Data processing was performed in the following steps. 
First, the Japanese tweets we collected were segmented into words using the Japanese morphological analysis tool Mecab~\cite{Kudo2005} with the Japanese dictionary NEologd~\cite{neologd}. 
Segmented tweets shorter than four words in length were considered less informative and deleted. 
As a result, there remained 312,169 tweets for training and 63,454 tweets for testing.
In this research, word2vec~\cite{Mikolov2013} was used as a word embedding method to create a dictionary of word vectors from the segmented tweets for training (30,5491 different words in 11,308,535 total words).
We used a Skip-gram model with a window size of 5 and 20 iteration times for word2vec implemented in the machine learning framework Chainer~\cite{chainer}.
Although doc2vec~\cite{Le2014} is often used for vectorization of sentences, it is unlikely to work for short sentences, such as tweets. 
Thus, instead of doc2vec, we used the method of averaging word vectors in order to obtain tweet vectors, as shown below (Fig.~\ref{pic1}B):
\begin{equation*}
\label{eq1}
T = \sum_{i=1}^{n}\frac{w_i}{n},
\end{equation*}
where $T$ is a tweet vector, $W_i$ is the vector of $i$ th word in a given tweet, and $n$ is the number of words in the tweet.

Tweets vectors were constructed based on data from 100 training accounts by referencing the dictionary of word vectors created previously (Fig.~\ref{pic1}D).
The same procedure was applied to data from 20 test accounts. 
Some words exist in tweets from 20 test accounts but did not exist in the dictionary of word vectors. 
We did not use these words for constructing tweet vectors. 
The number of unused words was only 4\% of the total words in the tweets from 20 test accounts.

Since a tweet consist of 140 characters or less, a single tweet may not convey enough personal information, but a collection of multiple tweets might be a more effective unit for inferring personal attributes.
Thus, we used a group of tweets or ``tweet blocks'' as inputs for machine learning and tweet block vectors were constructed by averaging the tweet vectors used (Fig.~\ref{pic1}C and D). 

\subsection{Machine Learning Algorithms}
Using machine learning algorithms, we trained and tested models based on single tweets ($L$=1) or either tweet blocks ($L > 1$) obtained from the above-mentioned processing method (Fig.~\ref{pic1}E). 
We used scikit-learn~\cite{scikit} for Linear Support Vector Classification (Linear SVC), K-Neighbors, AdaBoost, and Random Forest. 
The best parameters for these algorithms were selected using 10-fold cross-validation. 
Furthermore, we used Chainer for deep learning, in which a full connection neural network was chosen and parameters such as the number of middle layers, the number of nodes for each layer, learning rate, and the types of activation functions were optimized through repeated trials. 

\begin{figure*}[t]
\begin{center}
 \includegraphics[clip,width=1.0\textwidth]{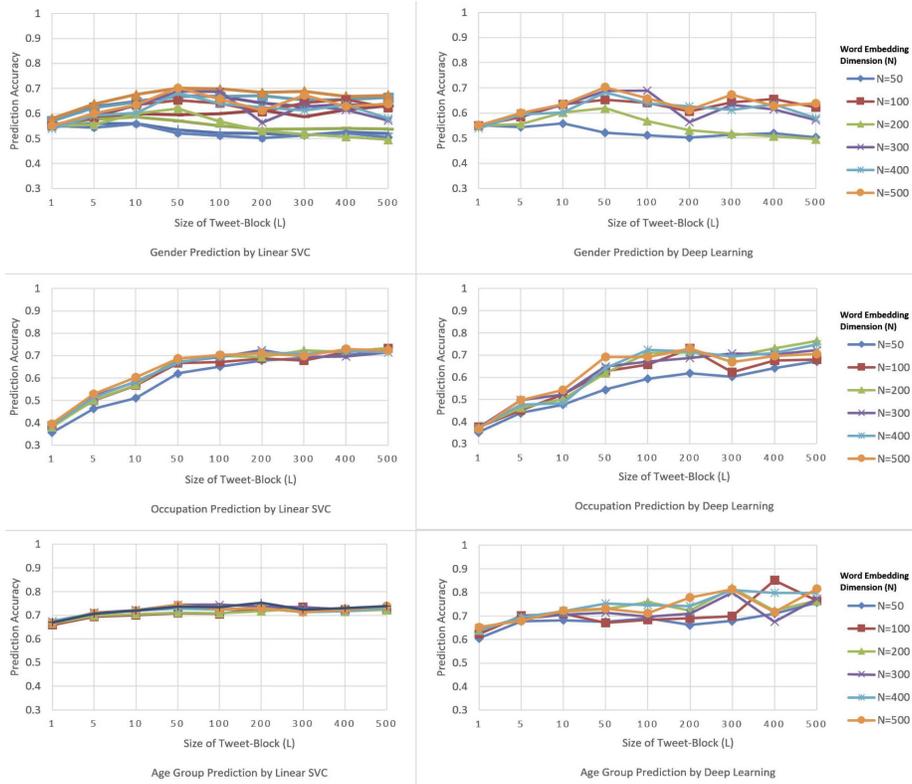}
\caption{Comparison of prediction accuracy between linear SVC and deep learning}
\label{pic3}
\end{center}
\end{figure*}

\begin{figure*}[h]
\begin{center}
 \includegraphics[clip,width=1.0\textwidth]{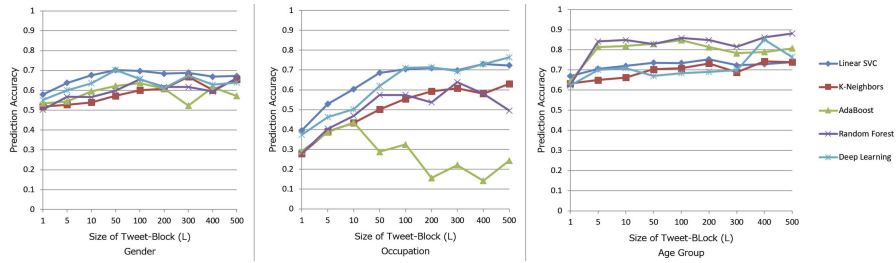}
\caption{Prediction accuracy as a function of $L$ under the condition where the best $N$ is selected for each attribute}
\label{pic4}
\end{center}
\end{figure*}

\begin{figure*}[h]
\begin{center}
 \includegraphics[clip,width=1.0\textwidth]{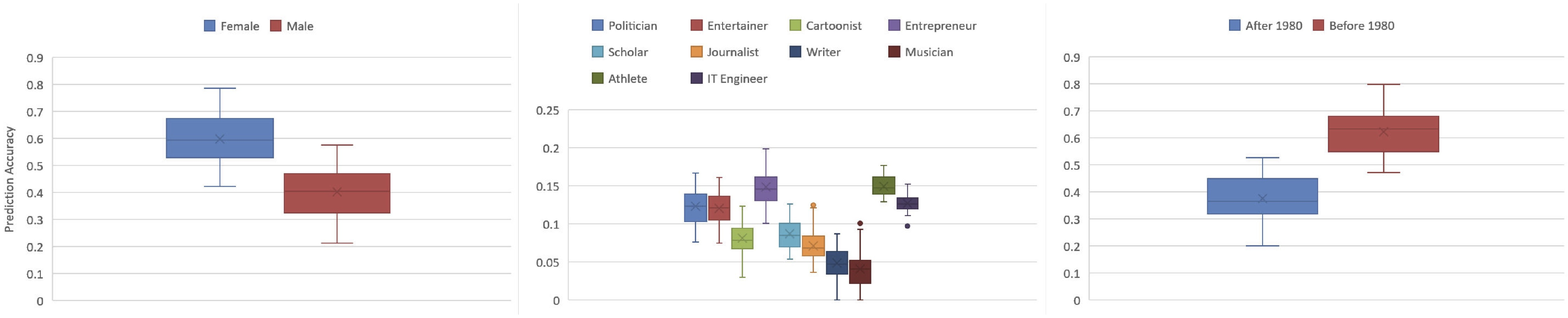}
\caption{Predictability within personal attributes}
\label{pic5}
\end{center}
\end{figure*}

\section{Results}
We examined the effects of the word embedding dimension ($N$) on word2vec, tweet block size ($L$), and different algorithms in terms of the prediction accuracy of three personal attributes: gender, occupation, and age groups. 

\subsection{Effects of Word Embedding Dimension and Tweet Block Size on Prediction Accuracy}
Figure~\ref{pic2} shows the accuracy of different learning algorithms for three kinds of prediction tasks in all combinations of $N$ and $L$. 
All the algorithms exhibited approximately 70\% accuracy with respect to inferring age groups for (digital natives and digital immigrants), indicating that age groups can be more easily predicted from social data than the other two attributes.  
The result also showed that Linear SVC and deep learning exhibit stable performance and higher accuracy compared with the two other algorithms.

Figure~\ref{pic3} shows the accuracy distributions of Linear SVC and deep learning for each attributes for different values of $N$ and $L$. 
For gender prediction, both Linear SVC and deep learning achieved approximately 70\% accuracy at $N$ = 500 and $L$ = 50 (F-Score = 0.688 and AUC-Score = 0.701 in Linear SVC; F-Score = 0.702 and AUC-Score = 0.706 in deep learning; male (positive) / female (negative)).
For occupation prediction, Linear SVC achieved approximately 70\% accuracy at $N$ = 500 and $L$ = 100, and  deep learning achieved a comparable level of accuracy at $N$ = 200 and $L$ = 100. 
For age group prediction, Linear SVC had a 75\% accuracy at $N$ = 500 and $L$ =200 (F-Score = 0.778 and AUC-Score = 0.749; digital immigrants (positive) / digital natives (negative)), and deep learning showed approximately about 80\% accuracy at $N$ = 200 and $L$ = 300 (F-Score = 0.821 and AUC-Score = 0.839; digital immigrants (positive) / digital natives (negative)). 
In most cases, a larger value for $L$ and $N$ led to a little higher accuracy, but this soon reached a plateau.

Figure~\ref{pic4} shows the prediction accuracy as a function of $L$ under the condition where the best $N$ was selected for each attribute.
For gender and occupation predictions, the best accuracy was achieved using Linear SVC and deep learning. In contrast, for age groups the best accuracy was achieved using Random Forest and AdaBoost. 
This result suggests that the optimal algorithm varies with the kinds of personal attributes, although Liner SVC and deep learning showed better and stable performance in our experiments.

\subsection{Differences in Predictability within Personal Attributes}
With regard to the three personal attributes, some entities have been thought to be more easily predictable than others. 
Intuitively, for example, politicians should be more predictable than others, since their word choice and usage would be unique. 
Figure~\ref{pic5} shows accuracy distributions for each attribute entities computed from all combinations of $N$ and $L$ in deep learning. 
In terms of the inference of personal attributes from texts, females were easier to predict than males ($t$-test, $P < 0.001$), entrepreneurs were easier to predict than musicians ($t$-test, $P < 0.001$), and digital immigrants were easier to predict than digital natives ($t$-test, $P < 0.001$).

\section{Discussion}
In this paper, we reported two preliminary results regarding the prediction of personal attributes (gender, jobs, and age groups) from text tweets by five different machine learning algorithms.

The results showed that the prediction of age groups is easier than the other two attributes. 
This can be explained by greater differences in word choice and usage between digital natives and digital immigrants, although this needs to validated in the future. 
All the algorithms exceeded 60\% accuracy for age group prediction, which was achieved even at $L$ = 1 (i.e., a single tweet) if selecting for a proper value of $N$. 
As for gender and occupation predictions, larger values of $L$ and $N$ were required to obtain more than 60\% accuracy ($N$ = 500 and $L$ = 50 for gender, $N$ = 500 and $L$ = 100 for occupation). Both predictions lead to approximately 70\% accuracy. 
It is encouraging that 60--70\% accuracy could be obtained for inferring personal attributes based only on text posts. 
This suggests that much higher accuracy could be achieved by adding other features, such as images or URLs embedded in tweets. 

In this study, deep learning did not show significant advantages compared with Linear SVC, but it has a potential advantage. 
While linear SVC needs to learn each attribute separately, deep neural networks (DNNs) can learn multiple attributes at the same time. 
If those attributes are mutually dependent, DNNs could learn faster with better prediction performance. 
This also needs to be tested in the future. 

Compared to single tweets, tweet blocks significantly improved accuracy for all prediction tasks, suggesting that the prediction of personal attributes require a certain number of tweets. 
In other words, tweet blocks as relatively low dimensional vectors can convey enough personal information such that personal attributes can be algorithmically inferred.  
According to our study, $L = 50$ is the lower limit for gender and occupation predictions of approximately 60\% accuracy. 
As shown in Figs.~\ref{pic2} and \ref{pic4}, even fewer tweets can yield a higher accuracy for the prediction of age groups.

Regarding the effects of the embedding dimension ($N$) for word2vec, better results were achieved in most cases when $N$ = 500 rather than $N$ = 50. 
In reality, however, greater $N$ requires larger computational costs. 
Thus, it is better to select an appropriate $N$ to balance computational costs and desired performance.

For future research, we will conduct a prediction test for other personal attributes by improving deep learning algorithms with a larger dataset. 
Once established a general framework for the inference of personal attributes can be applied to a wide variety of fields, including basic research on computational social science and applications for personalization and marketing.

\section*{Acknowledgment}
This research was supported by JST PRESTO Grant Number JPMJPR16D6, JST CREST Grant Number JPMJCR17A4, and JSPS/MEXT KAKENHI Grant Numbers JP15H03446 and JP17H06383 in \#4903. 


\end{document}